\documentclass[a4paper,12p]{article}
\usepackage{subfigure,graphicx,fullpage,natbib,footnote,times,wasysym,url,lineno}
\makesavenoteenv{tabular}\makesavenoteenv{table}

\newcommand{\halvepagina}{0.5\textwidth}

\newcommand{\fig}{figure~}
\newcommand{\Fig}{Figure~}
\newcommand{\citev}{\cite}
\newcommand{\Citev}{\citev}
\newcommand{\citeh}{\citep}
\newcommand{\eq}{equation~}
\newcommand{\tab}{table~}
\newcommand{\Tab}{Table~}

\newcommand{\Fair}{\F_\textrm{air}}
\newcommand{\Frolling}{\F_\textrm{rolling}}
\newcommand{\Power}{P}
\newcommand{\rrc}{C_R}
\newcommand{\F}{F}
\newcommand{\cw}{C_d}
\newcommand{\luchtdens}{\rho}
\newcommand{\opp}{A}
\newcommand{\spd}{v}
\newcommand{\mass}{m}
\newcommand{\energy}{E}
\newcommand{\dist}{s}
\newcommand{\Ukin}{\energy_\textrm{kin}}
\newcommand{\tm}{t}

\title{Electric Vehicles and Their Effect on Network Load}
\author{Victor L. Knoop}
\begin{document}
\maketitle
\section*{abstract}

The composition of the fleet of road cars is changing from fuel powered cars to plug-in hybrid (PHEV) and electrical vehicles (EV). The (electrical) range of these vehicles is limited, leading to so called ``range anxiety''. Firstly, this leads to a preference of shorter routes. Secondly, (PH)EVs have the ability to regenerate energy from braking, which make routes with many accerelations and decelerations not as unattractive for drivers of (PH)EVs as for drivers of vehicles with an internal combustion engine (ICE). This paper combines this, and adds drivers might prefer a lower speed limit due to a lower energy consumption. These elements are compared to  with road characteristics. It is found, motorways are usually less favourable for drivers of (PH)EVs compared to drivers of ICE vehicles, and they will prefer the shorter routes on secondary roads or through towns. 
 This will influence the use of the underlying road network, and thereby affecting congestion, emissions and safety. This shift towards the underlying road network needs to be taken into account in designs for the network. 

\section{Introduction}
Whereas the technical aspects of cars have improved over the past decades, their usage has been similar. Recently, electric vehicles (EVs) and plugin-hybrid electric vehicles (PHEVs) have been introduced. In these vehicles, drivers make use of battery power for propulsion. This has limitations on the range and speed of refuelling. Therefore, these vehicles can have an impact on the way people use the car: their way of driving and the route choice.

This is a recent development and hence the fleet does not consist of many vehicles which are (mainly) running on battery power. At 31 October 2014 43,059 (PH)EVs \citeh{nrEVs:2014},  were registered in the Netherlands, up from 30,112 at 1 Janaury 2014, a large increase in 10 months. At January 2014 129,673  cars were registered which had partial electric propulsion \citeh{hybrids:2014}, including (PH)EVs. Subtracting the number of (PH)EVs, we derive there are approximately 100,000 hybrid vehicles registered in the Netherlands. This is a relatively low number compared to the almost 8 million registered vehicles \citeh{nrEVs:2014}. However, the number of the EVs is very rapidly increasing, and this is simply not yet visible in the total fleet. The average age of a vehicle in the Netherlands is approximately 10 years \citeh{hybrids:2014}, suggesting that vehicles are taken of the roads typically 20 years after production. This is the time scale over which one can expect the EVs which are sold now to take effect on the traffic patterns.

The development of the road network takes place on a lower time scale, and takes at least decades to develop. This means that in network design one has to incorporate the effects which are starting to emerge. It has been studied earlier which effect the charging of EVs has on the charging structure (e.g., \citev{Neu:2014}), but the effect on the network loads has not been studied yet.
In this paper we will qualitatively consider the effect of the (PH)EVs on the network load, which can hence be used in network design or network management. 

The limited range of EVs is one of the main limitations for EVs and for PHEV which do not want to use the internal combustion engine (ICE). This could be for various reasons, for instance cost,  environmental, or ride quality. By consequence, the use of an EV is more limited, which has lead to various studies. For instance, the distances that are made \citeh{Lin:2012} can be compared to the range of a EV and the distance to a charging point. How often an EV is charged in practice, and the possible problems that users experience is shown by \citev{Tal:2014}.

\Citev{Nea:2013} performed a revealed preference study under the users of EVs. The most influential change for them compared to the experience with an ICE vehicle (ICEV) is ``the range and the battery drain.'' It is reported that trips are mostly made, but he reports on the experience of some drivers, in particular of worries on range and routes changes: ``Many drivers reported
however that they changed their driving style or their
chosen routes if they wanted to go on longer journeys: ‘If I
was on the rare occasion going on a long journey then
definitely I made the most of the regenerative braking and
tried to keep the battery life (..)
if it was a long journey then I
would definitely change my driving style just because I was
aware of the battery life.' ''
Their paper ends by showing that by choosing the right routes the range can increase.

Traffic assignment for EVs is shown by \citev{He:2013}. They show that the monetary costs per kilometer are lower for an EV than for a fuel powered car, but that the EVs are limited in range. The large consequences this has are shown for the Sioux Falls network. In the paper a constant cost per kilometer is assumed, for EV drivers lower than for ICEV drivers. Our paper extents this idea, showing possible effects on traffic assignment, but indicating differentiated kilometer costs based on the properties of the roads, leading to different energy consumption. We use the elements indicated in \citev{Nea:2013}, and add the effect of speeds which we compute from physical processes. 

The remainder of the paper is set up as follows. Section \ref{sec_litoverview} presents a background on car types and route choice. Then, section \ref{sec_valuationchange} shows the main changes in valuation of different aspects for drivers of EVs compared to drivers of an ICEV. Section \ref{sec_routetypes} briefly shows the characteristics of the routes, related to the aspects which have a changed valuation. Then, section \ref{sec_routes} shows the effects of the valuation changes on the routes. The paper ends with a discussion and conclusions in section \ref{sec_conclusions}

\section{Literature overview}\label{sec_litoverview}
This section first describes the different vehicle types and their characteristics. Section \ref{sec_lit_routechoice} then gives the basics of route choice modelling and traffic assignment.

\subsection{Car types}
In this section we differentiate between different cars, based on \citev{Tat:2009}. This section first explains the differences between the types (see also \tab \ref{tab_cartypes}).

The first is a traditional, fuel powered vehicle. Propulsion is generated only by an ICE. The second class is a hybrid vehicle. In this case, a electro motor can assist an internal combustion  engine in the propulsion. Some hybrids have the option for a complete electronic mode, where the electric motor is operating, but the internal combustion engine is not running. Essential to a hybrid is that only fuel goes into the vehicle. Ultimately, the charge for the batteries hence have to come from the fuel. In some cases, the internal combustion engine can charge the batteries. The other option is that the batteries of the electro motor are charged by the driving, if the driver reduces speed more than the resistance. This is called regenerative braking. This regenerative braking is also present in the other vehicles classes with an electro motor.

In the third category, the plug-in hybrid, the same propulsion concept holds as for the regular hybrid, but the batteries can be charged with external electrical power. Electrical ranges differ highly (see also table \ref{tab_carref} later on). The fourth category of cars are electric vehicles (EVs). The only propulsion in an EV is the electrical motor. They are usually charged with electrical power from the grid. Depending on the use and expected charge, different usage of the fuel engine and battery power can be selected \citeh{Rog:2013}.

At this moment, the typical maximum distance for an EV on a fully loaded battery is approximately 150 km. Charging the batteries on the grid takes a long time, and hence drivers are afraid they can run out of energy, which is now mentioned as ``key barrier'' \citeh{Nea:2013} for the adaptation of EVs. This is commonly referred to as ``range anxiety'' (see e.g., \citeh{Neu:2014}). This is overcome by a fifth category of cars, the EV with range extender. This is an EV with a internal combustion engine that runs on fuel and can charge the batteries. Note that contrary to the plug-in hybrid the in the EV with range extender there is no direct connection from the internal combustion engine to the wheels. In EVs, the electrical motor has to be more powerful than in plug-in hybrids since it is the only power source available.

\begin{table}\caption{Properties of vehicle types -- typical values}\label{tab_cartypes}\centering
\begin{tabular}{l|llll|llll}\hline
&ICE&ICE power&Electro&Grid&\multicolumn{4}{c}{Consumption/cost for a 100 km trip}\\
&    &to wheels& motor&power&Range&Fuel&Electricity&Costs\\
&    &         &      &     &(km)      &(liter)     &(kWh)            &(eur)\\\hline
Internal combustion&$\bullet$ &$\bullet$	&&&500 +&8&&14\\
Hybrid&$\bullet$	&$\bullet$	 &$\bullet$ &&500 +&5&&8,75\\
Plug-in hybrid&$\bullet$ &$\bullet$ &$\bullet$ &$\bullet$ &500 +&4&4&8\\
EV with range extender&$\bullet$ &&$\bullet$ &$\bullet$ &300&&20&5\\
EV&&&$\bullet$ &$\bullet$ &150&&18&4,5\\\hline\end{tabular}\end{table}

The right hand side of \tab \ref{tab_cartypes} shows the different car types and some typical usage costs. Costs for fuel and electricity vary per country. For the table, typical Dutch prices of 1.75 euro/liter unleaded fuel and 0.25 euro/kWh for electricity. The consumption also differs within the same type of vehicles, but that is neglected for this overview. Moreover, usage costs differ for the different road types, which we will discuss in detail (qualitatively) in section \ref{sec_valuationchange}. 

Earlier work \citeh{He:2013} combines the effects of the range anxiety for EV drivers with the lower cost per unit of distance. This has shown to have a large effect on the assignment. However, that paper does not account for differences in usage cost per unit of distance for the different road types. This paper explicitly looks into these effects.

\subsection{Route choice modelling and traffic assignment}\label{sec_lit_routechoice}
Most route choice models are based on utility. That means that each route has characteristics and the joint of characteristics is valued by a user by a utility. There are many variations using this basis, allowing for random variations in the (interpretation of the) characteristics, of random (or personal) variations in valuation of the characteristics. The choice can be made for the alternative with the highest utility, or under uncertainty the alternative with the lowest possible regret.

Any of these methods use the basis of a utility function which maps the various characteristics into a utility. In this paper we will consider the partial valuation of three characteristics, being (i) speed, (ii) distance and (iii) accelerations (i.e., changes in speed). It is discusses how these differ for drivers of different types of cars (section \ref{sec_valuationchange}). 

The most common approach to traffic assignment is a so called equilibrium assignment. This means that traffic is in the state of Wardrop equilibrium \citeh{War:1952}, meaning that no vehicle can change its route unilaterally and improve its travel cost. Note that this equilibrium is a user equilibrium. In his famous paper \citev{Bra:1968} shows the difference with the system optimum, in which every one can have a shorter travel time than in the user optimum

Most common costs to include in the route costs are time, which can be translated into monetary cost by a value of time, and monetary costs (for driving or explicitly tolls). Some of the costs are stochastic, and should be considered for all drivers individually. This extension is first shown by \citev{Dag:1977}. For more information we refer to \citeh{She:1985}.  
\citev{Bel:1995} adds queuing to this principle. Later, solutions for different user preferences and multi-criterion objectives are presented (e.g., \citev{Che:2011EVS,Zha:2013EVS}). If cost functions are determined, these methods can be used to solve the assignment problem.

\section{Expected changes in valuation}\label{sec_valuationchange}
This section describes how drivers are expected to have a different valuation of characteristics of a road depending on their vehicle type. This is based on drivers reporting on their behaviour \citeh{Nea:2013}, as well as technical details of the vehicles and the resulting physical possibilities and limitation. Some of the calculations, especially in section \ref{sec_speedvaluation} and \ref{sec_acccost}, may be more complicated than the average (PH)EV driver is aware. However, the (PH)EVs usually are equipped with screens informing the driver on the energy consumptions and levels. Therefore, drivers are can derive the consequences of particular driving behavior themselves.

\subsection{Valuation of distance}\label{sec_distancevaluation}
A driver worried on his range \citeh{Nea:2013}  is likely to choose the shortest route from his origin to his destination. Note that this is separate from the fuel efficiency, discussed in the next sections: also if the miles per gallon were to remain constant, the drivers would look for a short route. Hence, the costs per unit distance will be higher for an EV user.

\subsection{Valuation of speed}\label{sec_speedvaluation}
There are two main resistances, the rolling resistance and the air resistance. The rolling resistance ($\Frolling$) is commonly expressed as 
\begin{equation}
\Frolling=\rrc \mass g\label{eq_frolling}
\end{equation}
In this equation $\rrc$ is the rolling resistance coefficient of the tire, and $\mass$ the mass of the vehicle, and g the gravitational acceleration of 9.81 m/s$^2$. Note that this rolling resistance force is not dependent on the speed of the vehicle. Modern eco-tires optimized are optimized for low rolling resistance and have a $\rrc$ of 0.006 \citeh{RolRes:2006}. This means that the resistance force is 6 {\permil} of the normal force acting upon the vehicle. 

Characteristics of various vehicles are listed in \tab \ref{tab_carref}. Throughout this section, several calculations are made in which a Nissan Leaf is taken as example. Using \eq \ref{eq_frolling}, a rolling coefficient of 0.006, and mass of 1525 kg is 90 N. 

The air resistance $\Fair$ is expressed as follows
\begin{equation}
\Fair=\frac{1}{2}\cw \luchtdens \opp \spd^2\label{eq_fair}
\end{equation}
In this equation, $\luchtdens$ is the density of air (1,3 km/m$^3$), $\opp$ the frontal area of the car, $\spd$ the speed of the car and $\cw$ the drag coefficient. Using the values of \tab \ref{tab_carref}, we find the resistance force for a Nissan Leaf travelling at 130 km/h (36 m/s) of 610 N.

\begin{table}
\caption{Characteristics of some hybrid and electrical vehicles (2014 model year)}\label{tab_carref}
\small
\begin{tabular}{lllllll|ll}\hline
Car&Type&Weight&Electric&Cd&\opp (m$^2$)&Cd\opp&Regen.&Stop\\
&&(kg)&power (kW)&&(m$^2$)&(m$^2$)&frac. (\%)&dist (m)\\\hline
Honda Civic Hybrid&Hybrid&1303&17&0.27&1.87&0.50&82&580\\
Toyota Prius &Hybrid&1380&60&0.25&2.17&0.54&93&210\\
Toyota Prius plug-in hybrid&Plug-in Hybrid&1436&60&0.25&2.17&0.54&93&210\\
Volvo V60 D6&Plug-in Hybrid&2000\footnote{Official claim is 1961-2066 depending on equipment}&50\footnote{50 kW temporarily, 20kW continuously}&0.29&2.28&0.66&90&340\\
Chevrolet Volt&Plug-in Hybrid/EV\footnote{Technically a plug-in hybrid (driveshaft from combusion engine to the wheels), in practise an EV since the driving is set-up for electrical driving}&1721&111&0.28&2.22&0.62&96&140\\
BMW i3&EV&1195&125&0.29&2.38&0.69&96&90\\
BMW i3&EV + range ext&1315&125&0.29&2.38&0.69&96&100\\
Nissan Leaf&EV&1525&80&0.28&2.59&0.72&93&170\\
Tesla Model S&EV&2108&310&0.25&2.30&0.58&98&70\\\hline
\end{tabular}
\normalfont
\end{table}

The power $\Power$ required to overcome these forces can be calculated using 
\begin{equation}
\Power=\F*\spd\label{eq_powercalc}
\end{equation}
The total amount of energy needed ($\energy$) is the integral of the power $\Power$ over time.
\begin{equation}
\energy=\int \Power dt\label{eq_powertime_int}
\end{equation}
Note that for constant power during a time interval $\tm$ the energy can be expressed as
\begin{equation}
\energy=\Power \tm\label{eq_powertime}
\end{equation}
The time interval $\tm$ needed to travel a given distance $\dist$ at a speed $\spd$ is given by $\tm=\dist/\spd$. Hence, substituting this into \eq \ref{eq_powertime} we find an energy consumption
\begin{equation}
\energy=\Power \frac{\dist}{\spd} = \F*\spd\frac{\dist}{\spd} = \F \dist \label{eq_energypropforce}
\end{equation}
This shows that the power needed for a road stretch travelled at constant speed is proportional to the force needed to overcome the resistance at that speed. 

\begin{figure}
\subfigure[Regenerative braking]{\includegraphics[width=\halvepagina]{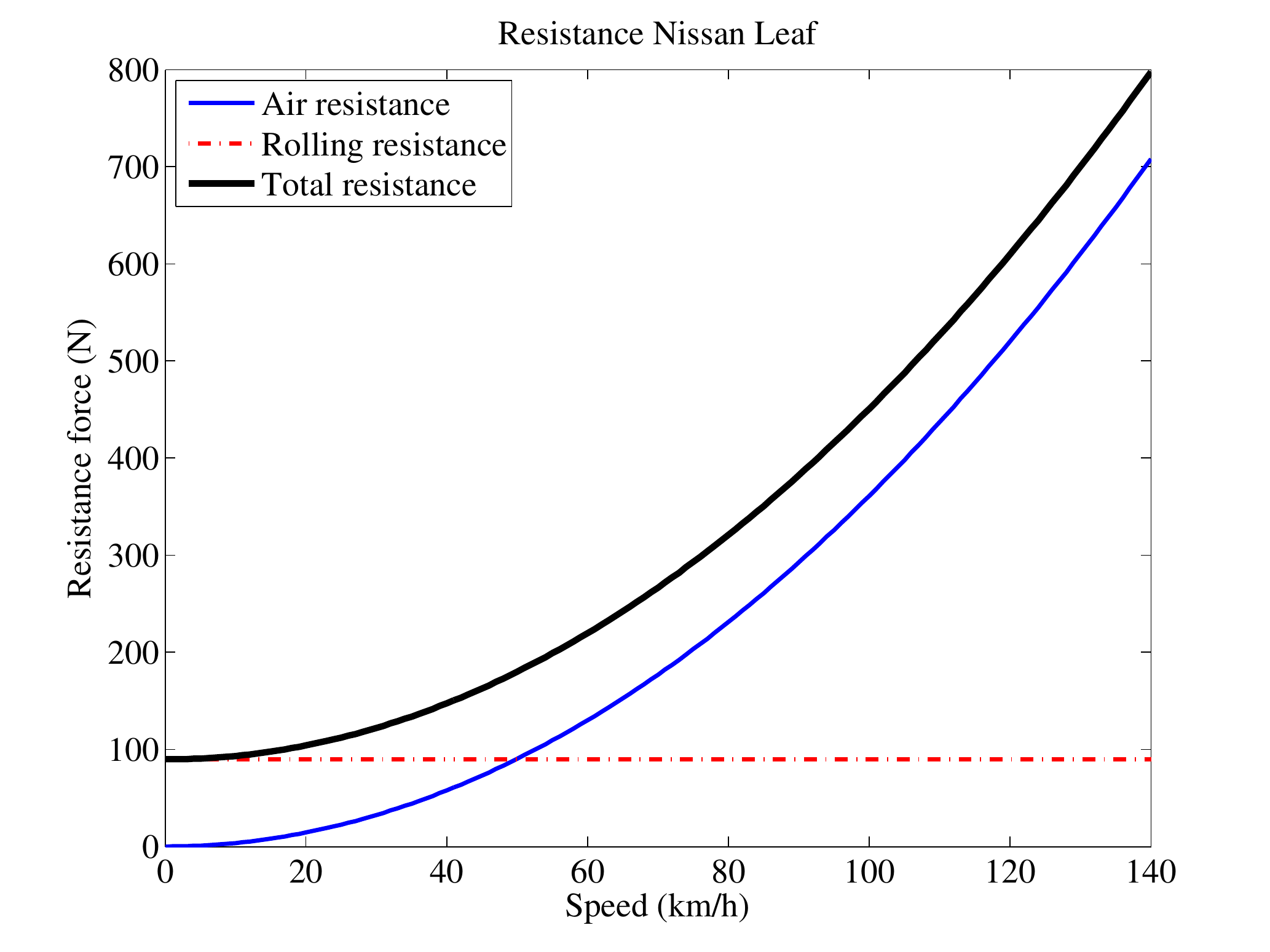}\label{fig_resistance_leaf}}
\subfigure[Regenerative braking]{\includegraphics[width=\halvepagina]{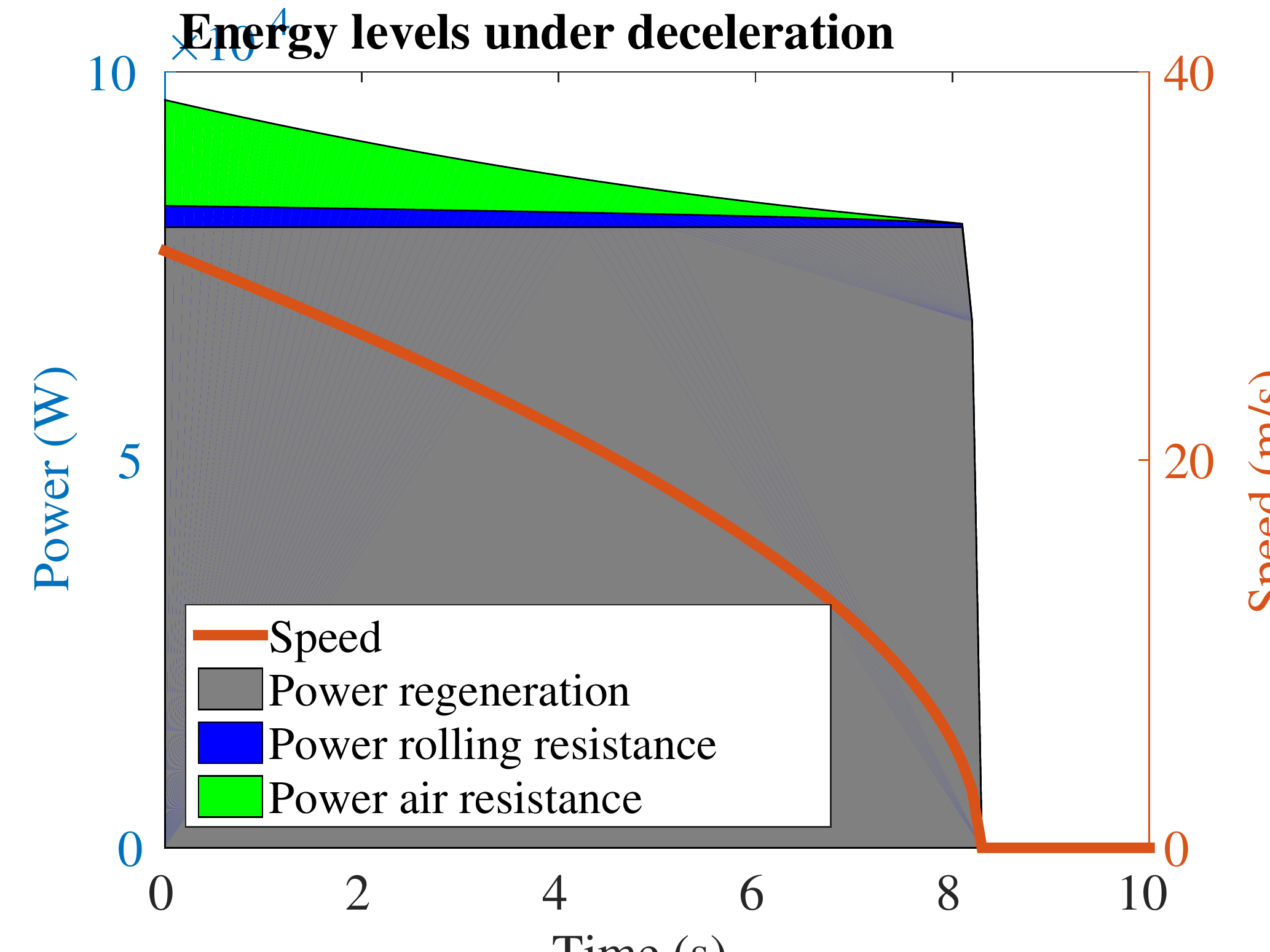}\label{fig_decleration_pattern}}
\caption{Energy management -- illustration for a Nissan Leaf}
\end{figure}

\Fig \ref{fig_resistance_leaf} shows how the resistance forces change as function of speed. The air resistance increases quadratically with speed, and for speeds over urban speeds becomes the dominant factor for energy consumption.
Note that for any car, a reduction from 130 km/h to 80 km/h would be a reduction of air resistance by a factor of $(130/80)^2=2.6$. For the Nissan Leaf this is a reduction from 610 N to 231 N. 
The total amount of energy needed to travel at 80 km/h can be expressed as a fraction of the energy needed to travel at 130 km/h. Since the energy consumption is proportional to the force (equation \ref{eq_energypropforce}), this is (231+90)/(610+90) = 46\% of the original energy consumption (for a Nissan Leaf), or less then half the energy consumption. Hence, drivers of an EV worried about energy are expected to assign a high cost to driving at high speeds.

\subsection{Valuation of acceleration}\label{sec_acccost}
For ICEVs, acceleration is a main source of fuel consumption. The kinetic energy of a moving object ($\Ukin$) is calculated from its speed $\spd$ and mass $\mass$, using the equation:
\begin{equation}
\Ukin=\frac{1}{2}\mass\spd^2\label{eq_kinenergy}
\end{equation}
For the Nissan Leaf, the kinetic energy for travelling at 80 km/h (22 m/s) is 377 kJ. 

This value can be compared to the energy consumption rate for constant speed, calculated by \eq \ref{eq_powercalc}. For the Nissan Leaf the required force is 231+90=321 N, or .32 kN. Translating this into power, using \eq \ref{eq_powercalc} gives a power consumption of
\begin{equation}
\Power=\F*\spd=.321 * 22 =7.5 \textrm{kW}\label{eq_airres}
\end{equation}
The time a certain amount of power can be generated with a given amount of energy, can be calculated using (the inverse of) \eq \ref{eq_powertime}. For the same amount of energy that the car at 80 km/h has a kinetic energy, it can overcome the resistance forces at 80 km/h during a time $\tm$ = 377kJ/7.5kW = 50 seconds. At this speed, this equals $\spd \tm$ = 80/3.6 x 50 = 1.1 km of driving. 
Accelerating, for instance after a full stop at a traffic light, will hence consume a large part of the energy level of the car, and reduces the total range by 1.1 km. Note that this is the reduction of total distance which can be travelled on a charge reduces by 1.1 km, apart from the fact that the remaining distance reduces because during acceleration and braking one covers distance.

In most ICEVs, kinetic energy is transferred into heat while braking. So each stop will cost much energy. This is not the case for electrical vehicles. The electrical motors can operate in the reverse direction, by which the kinetic energy is regenerated and transferred into electrical energy, which can be reused for acceleration. Therefore, in EVs, the energy consumption will not drastically increase at secondary roads with stops. Hence, they are more viable as alternative.

Figure \ref{fig_decleration_pattern} shows the deceleration pattern of a Nissan Leaf from 80 km/h to standstill under maximum regenerative braking (80 kW, see table \ref{tab_carref}). The height of the blocks is the power, and the total amount of energy is the integral of power over time (equation \ref{eq_powertime_int}), which hence is represented by the shaded area. The power dissipated by the rolling resistance and the air resistance are calculated by their respective forces (\eq \ref{eq_frolling} and \eq \ref{eq_fair}), substituted in \eq \ref{eq_powercalc}. The kinetic energy reduced sum of the energy dissipation by the rolling resitance, the energy dissipation by the air resistance and the energy regeneration by the electo motor, leading to a new kinetic energy and a new speed (\eq \ref{eq_kinenergy}).

The reduction of energy can be split into different sources, being either regenerative braking or resistance. The area of the blocks shows which amount of energy is regenerated and which amount is lost due to resistance. Comparing the areas shows that in case of the Nissan Leaf, 93\% of the energy is regenerated under a braking distance of 170m. \Tab \ref{tab_carref} lists the regeneration fractions and stopping distances for various vehicles under maximum regenerative braking. If the braking distance needs to be shorter (for instance in emergencies), kinetic energy needs can be dissipated in terms of heat by conventional braking.

\subsection{Generalisability}
In summary of section \ref{sec_distancevaluation}-\ref{sec_acccost}, drivers of EVs prefer shorter distances, lower speeds and accelerations form a lower cost compared to drivers of ICE powered vehicles.  To some extent, the remarks on energy consumption and speed and range as made in sections \ref{sec_distancevaluation} and \ref{sec_speedvaluation} also hold for drivers of ICEVs. 
However, users of EVs are expected to be more responsive to those facts, due for two reasons: (i) the range is much more limited (ii) there are no options to quickly refuel (iii) the are directly confronted with their actions by the range indicator.  That means that also with a strongly increased range this effects might hold. Other potential future effects like a strong increase in the energy prices can cause similar effect for ICEV drivers. 

ICEVs have a strong disadvantage of acceleration, as shown section \ref{sec_acccost}. That means that even under strongly increased fuel costs, a short route with many accelerations (traffic lights, roundabouts) will not become attractive -- it might even become less attractive. Hybrid cars can store the energy that becomes available under braking to some extent. This regenerative process is generally less strong than with full electric vehicles. Hybrid vehicles can therefore be considered as an in-between vehicle class. Even within hybrid vehicles, there are differences. Mild hybrid vehicles have a electrical motor with limited power and hence the regenerative power is equally limited. The estimated effects on the valuation for different characteristics are summarized in the left hand side of \tab \ref{tab_utilities}.

\begin{table}
\caption{The expected sensitivity of utility for characteristics and the characteristics of road types}\label{tab_utilities}
\begin{tabular}{lcccc|ccc}
\hline 
&\multicolumn{4}{c|}{Utility for...}&\multicolumn{3}{c}{Characteristics of...}\\
&ICE&Hybrid&Plug-in hybrid&EV&Motorway&Secundary&Urban\\\hline
Speed&-&-&- -&- - - (for high speeds)&+++&+&- -\\
Distance&-&-&- -&- - -&++&0&- -\\
Acceleration&- -&-&-&0&- -&+&+ + +\\\hline
\end{tabular}
\end{table}

\section{Type of routes}\label{sec_routetypes}
In this paper, we differentiate between motorways, secondary roads and urban roads. Motorway are multi-lane roads with a high speed limit (typically 100-130 km/h). Drivers can maintain their speed since there are no interruptions by junctions or traffic lights. In urbanized areas, the motorways usually make a detour around the cities. 

Secondary roads are road with a speed limit of typically 70-90 km/h. There are traffic lights which make the traffic stop for crossing traffic. The roads are not necessarily multi-lane. Secondary roads are often found in the surroundings of urbanised areas, allowing access to the country side, but thereby possibly connecting different parts of the network. 

The third type of roads are urban roads. Speed limits are low, usually around 50-60 km/h. Traffic needs to stop frequently for traffic lights and speeds are highly volatile.  An overview of these characteristics is given in right hand side of \tab \ref{tab_utilities}.

\section{Effects on the route choice}\label{sec_routes}
\begin{figure}
\includegraphics[width=0.75\textwidth]
{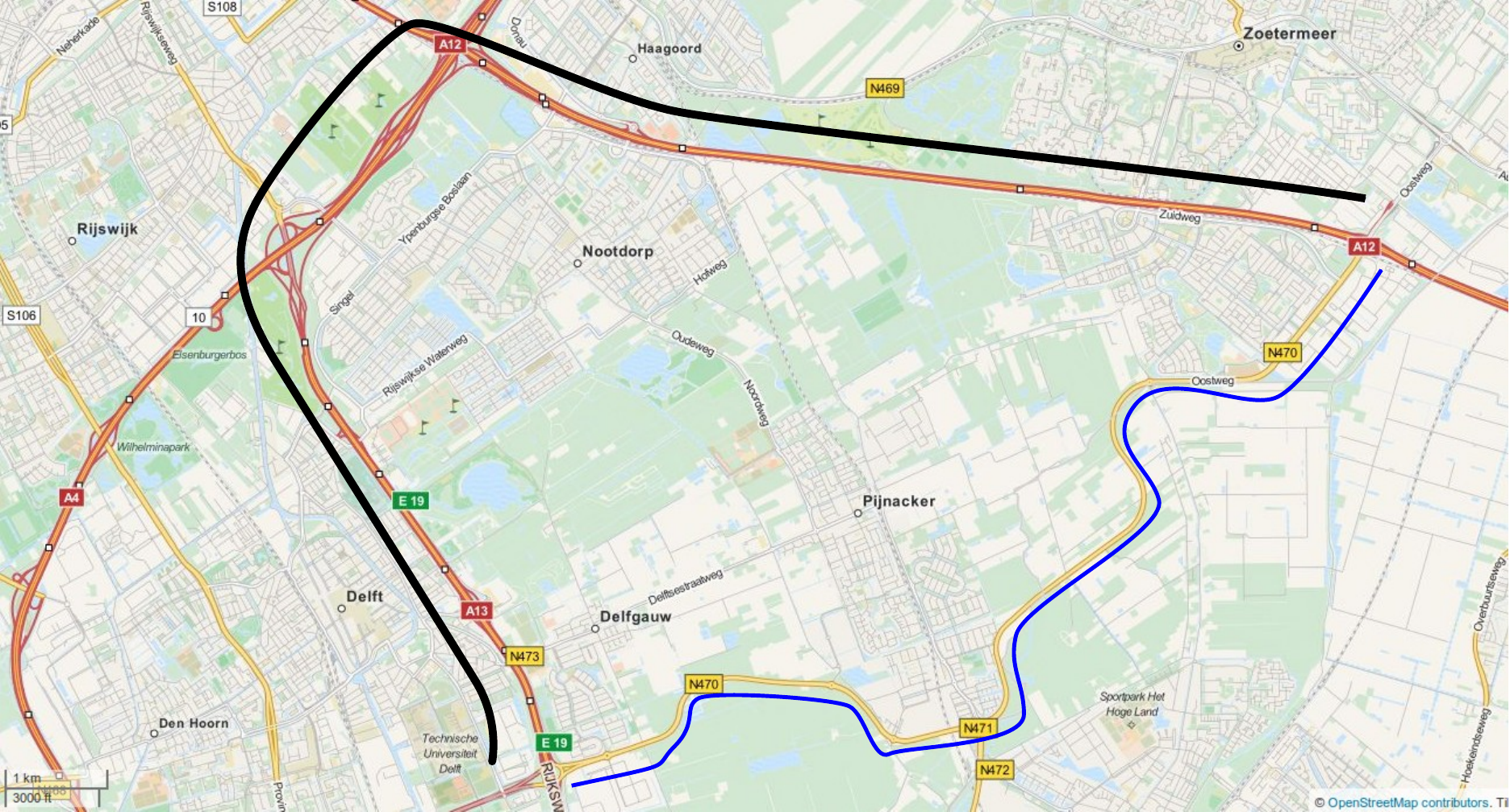}
\caption{Two roads which can be taken}\label{fig_map}
\end{figure}
A typical trade-off in current networks is a choice between a shorter trip on a secondary road, or a longer trip on a freeway. These choices are more or less equally valued since the travel time is considered leading for the trip choice. There is also a geographical reason why these trip occur. In spatial planning, the main roads will often go through the less populated area. There might be ring roads around the populated area, and secondary roads through the area. Due to the road type, the travel time on the road around the urban area can be similar. In fact, only if this travel time is comparable, the road is constructed.

Insight of this paper is that, combining the valuations in the left side of \tab \ref{tab_utilities} with the route characteristics in the right side of of \tab \ref{tab_utilities}, drivers of EVs will make a different choice. They assign higher costs to distances, to high speeds, both disfavoring the ringroad. Moreover, added utility for the ringroad due to lower accelerations is lower for EVs, shifting preference even further away from the ringroad. Without other measures, this will mean that traffic (EVs) will shift from the ring roads to secondary roads. 

In a real world traffic situation, the traffic will consist of different users all having their own valuation. This could be described in classes of similar valuation, e.g., (i) fuel powered cars, (ii) hybrids, (iii) PHEV, (iv) EVs. A multi-class traffic assigment (e.g., \citev{Daf:1972}) could be carried out, where choices of all classes will influence the traffic state on a link, but each vehicle class will optimize its own objective. In other words, all classes have the same traffic operations, but they value the different characteristics differently, and hence they choose different routes. 

This can have a large effect on the traffic volumes though the urbanised areas. That, in turn, has possible large consequences. The main effects are \begin{enumerate}
\item{Travel time increase;}
\item{Increased interactions with the traffic on the local roads, both motorized and non-motorized;}
\item{Safety issues, since the motorways are safer per distance travelled than secondary roads;}
\item{Possibly noise issues, since although the EVs have are silent motors, at speed the tires will produce a noise. Secondary roads are usually closer to living areas.}
\end{enumerate}

For instance, consider the trip from Zoetermeer to Delft, depicted in \fig \ref{fig_map}. One can either take the longer freeway route (depicted in blue, 19.5 km, 13 min in free flow) or the shorter route using the secondary road (depicted in black, 13 km, 14 min in free flow). This situation occurs very frequent, since often there are ring roads around more urbanized area for the though traffic, but there are also roads through the more urbanised area.
Traditional user equilibrium on travel times will load both routes to flow where travel times are equal. The new insights will give users of EVs an incentive to take the shorter route, possibly at the cost of a higher travel time. That means that the load on the secondary road will be larger, leading to the 4 possible consequences mentioned above.

Earlier work \citeh{He:2013} showed that there is a difference in driving costs per mile for EVs and fuel powered cars. The limited range would limit the use of long routes, but the monetary cost per mile was considered lower for EVs than for fuel powered cars, due to the difference in price between fuel and electricity (0.04 vs 0.13 euro per mile was assumed). The arguments in this paper indicate that drivers in EVs are likely to assign a higher cost to long and fast routes, rather than lower.

\section{Discussion and conclusions}\label{sec_conclusions}
This paper shows that route choices might change with changing fleet composition. In particular a larger share of (PH)EVs will move the traffic equilibrium from the motorways surrounding the urban areas towards straighter routes through the urbanised areas.  

It goes beyond the scope of this paper to asses all these factors qualitatively and provide a road design. The main message of the study here is that the default settings for the cost function of drivers in the network equilibrium assignment should be changed. As argued, these cost functions are user class specific, and a class specific traffic assignment can be made.

For a new road design travel time is only part of the decision variables. Other variables include noise, pollution and safety. These effects change due to the changed routes, and should also be considered class specific. For instance compared to fuel cars, EVs produce less noise and are less polluting. 

The time scale of the transition of route preference is shorter than the average age of the routes. This is a transition which takes place at a time scale of approximately 10 years, the average age of a vehicle. The road network changes more slowly. Therefore, changes in the fleet affecting the road use should be addressed now in order to be aligned with the changing fleet.

Road designers should account for new preferences. They can either try to accept the new preferences and adapt the road network, or interfere in the choice process in order to keep the current route choice distribution. If they choose to accept, they need to 
adapt the roads to the new equilibrium. 
This can imply that the attention for road improvement should lie in improving the direct routes, rather than increasing the capacity of the current motorways. 

Alternatively, the can try to interfere. A first way is use ``traditional'' method of speed limit adaptation. A lower speed limit in the urban area can make the direct route across town less attractive. These could be measures to reduce speed by speed limits or traffic lights. 

An alternative solution is to apply pricing in such a way that the tolls account for the marginal effects on the system. This could be done with equal tolls for all classes, thereby optimizing the traffic volumes. Alternatively, one can have a class-specific toll. This way, the system can take advantage of the qualities of each the specific vehicle classes. For a discussion on the possibilities of tolls, see \citev{Yan:2004EVS}.

Independent of the road changes, car manufacturers advice their drivers on the best route to choose. Although the manufacturers are sparse with information due to reasons of competitiveness, some of the route preferences mentioned here might already be implemented in in-car navigation devices. Also providers of plug-and-play satellite navigation systems or navigation apps on smart phones can apply a setting specially focussed on EVs.  They can be adjusted such that the chosen route depends on the remaining energy and the expectation where the vehicle can be charged.

\bibliographystyle{trailthesis}
\bibliography{arXivpaper}
\end{document}